\documentclass[12pt,a4paper]{article}
\usepackage{epsfig}
\usepackage{graphicx}                  % provides \includegraphics command
\usepackage{ae}
\usepackage{amsmath}
\usepackage{amssymb}
\usepackage{graphics}
\usepackage{dcolumn}
\usepackage{bm}
\begin{document}
\begin{center}

{\bf \Large ALICE TPC upgrade for High-Rate operations}\\
\vspace{0.1cm}
{\bf  Saikat Biswas On behalf of the ALICE Collaboration}\\
\vspace{0.1cm}
{\bf School of Physical Sciences, National Institute of Science Education and Research, Jatni-752050, India}\\
\vspace{0.1cm}
{\bf E-mail: saikat.ino@gmail.com, s.biswas@niser.ac.in, saikat.biswas@cern.ch}

\end{center}
%\ShortTitle{ALICE TPC upgrade for High-Rate operations}

%\author{\speaker{Saikat Biswas}\thanks{On behalf of the ALICE Collaboration}\\
       % School of Physical Sciences, National Institute of Science Education and Research, Jatni-752050, India\\
       % E-mail: \email{saikat.ino@gmail.com, s.biswas@niser.ac.in, saikat.biswas@cern.ch}}

%\author{Another Author\\
%        Affiliation\\
%        E-mail: \email{...}}

\abstract{A new type of Time Projection Chamber (TPC) has been proposed for the upgrade of the ALICE (A Large Ion Collider Experiment at CERN) so as to cater to the high luminosity environment expected at the Large Hadron Collider (LHC) facility in future. This device will rely on the intrinsic ion back flow (IBF) suppression of Micro-Pattern Gas Detectors (MPGD) based technology in particular the Gas Electron Multiplier (GEM). GEM is to minimise the space charge effect in the main drift volume and thus will not require the standard gating grid and the resulting intrinsic dead time. It will thus be possible to read all minimum bias Pb--Pb events that the Large Hadron Collider (LHC) will deliver at the anticipated peak interaction rate of 50 kHz for the high luminosity heavy-ion era in Run 3. New read-out electronics will send the continuous data stream to a new online farm at rates up to 1~TByte/s. The new read-out chambers will consist of stacks of 4 GEM foils combining different hole pitches. In addition to a low ion back flow ($<$ 1\%) other important requirements are good energy resolution (better than 12\% (sigma) for $^{55}$Fe X-rays) and operational stability.}

%\FullConference{7th International Conference on Physics and Astrophysics of Quark Gluon Plasma,\\
		%1-5 February , 2015 \\
		%Kolkata, India }

%\begin{document}

\section{Introduction}

The ALICE Experiment at the LHC facility at CERN is upgrading the central barrel detectors to cope with the foreseen increase of the LHC luminosity in Pb--Pb collisions after 2018/19 \cite{ALICE}. The main goal of this upgrade is to record Pb--Pb interactions at a rate of 50 kHz after the Long Shutdown 2 (LS2) in the year 2018-2019, which is a factor of about 100 times more than the current data acquisition rate of about 500~Hz. To overcome the rate limitations imposed by the present gated read-out scheme,  the existing Multi Wire Proportional Chamber (MWPC) based read-out chambers will be replaced by GEM based detectors for the TPC \cite{FS97}. 

In this article a brief overview of the R\&D effort on GEM based detectors, test results obtained in India and elsewhere for the ALICE TPC upgrade is discussed.
	
\section{The ALICE TPC: Present and future}
The ALICE TPC is the largest gaseous Time Projection Chamber in the world with an active gas volume of about 90~m$^3$. The TPC (as shown in Figure~\ref{tpc}) has a cylindrical field cage with a central high voltage electrode and a read-out plane on each endplate, presently consisting of 72~Multi Wire Proportional Chambers (MWPC), with a total of about 550000 read-out cathode pads. The operating gas mixture is Ne/CO$_2$ (90/10) or Ar/CO$_2$ (90/10) in Run1 and Run2 respectively \cite{ALICEtpc}. The present MWPC read-out chambers are operated with an active Gating Grid (GG). To reduce the event pile-up in the 50~kHz Pb--Pb running scenario in Run3, the TPC will have to be operated continuously and the back flow of ions needs to be minimized without the use of a GG.

%%%%%%%%%%%%%%%%%%%%%%%%%%%%%%%%%%%%%%%%%%%%%%%%%%%%%%%%%%%%%%%%%%%
\begin{figure}[htb!]
\begin{center}
\includegraphics[scale=0.4]{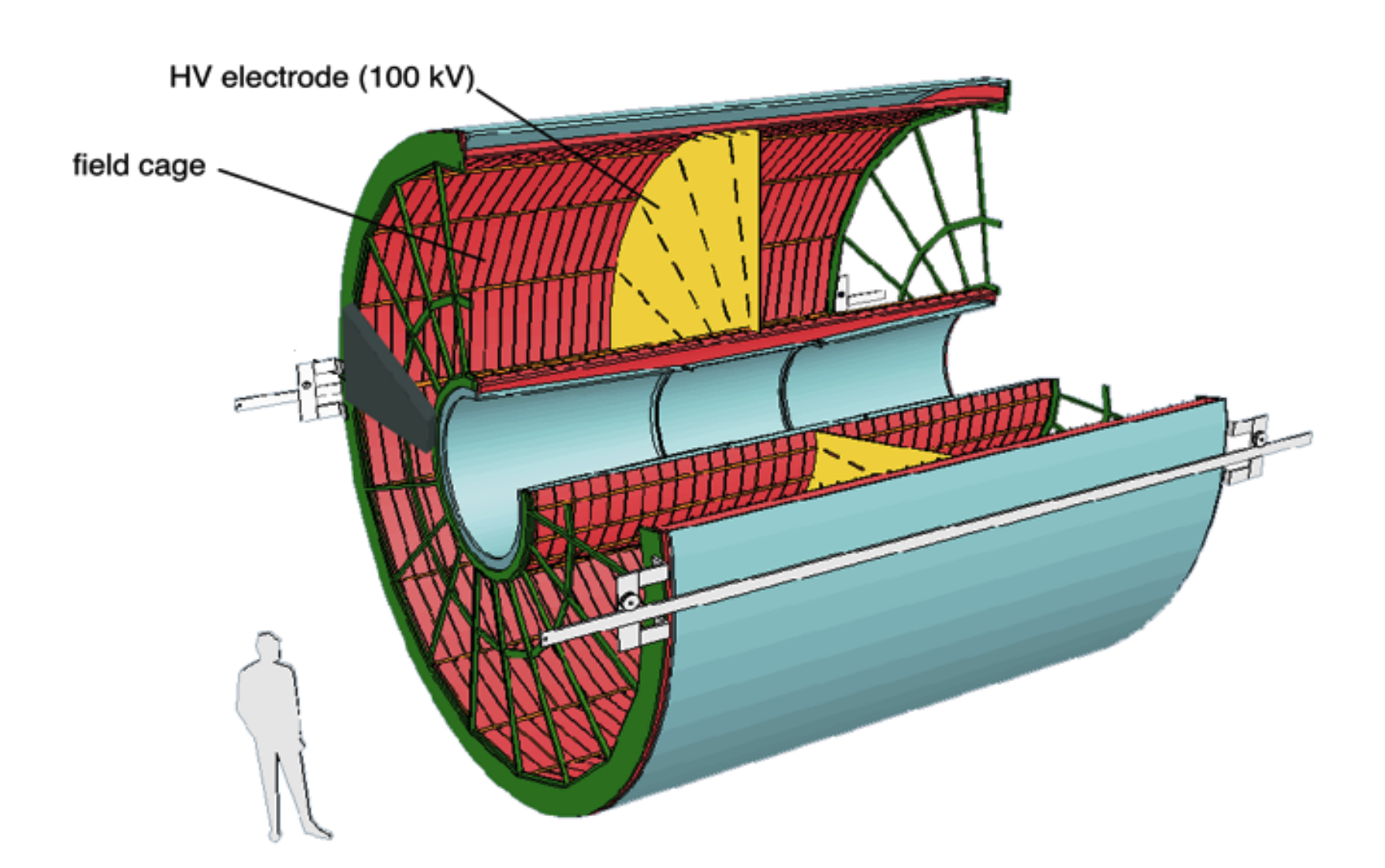}
\caption{\label{tpc}Schematic view of the ALICE TPC. The current MWPC-based read-out chambers will be replaced by 4-GEM stacks \cite{ALICE}.}\label{tpc}
\end{center}
\end{figure}
%%%%%%%%%%%%%%%%%%%%%%%%%%%%%%%%%%%%%%%%%%%%%%%%%%%%%%%%%%%%%%%%%%%
%%%%%%%%%%%%%%%%%%%%%%%%%%%%%%%%%%%%%%%%%%%%%%%%%%%%%%%%%%%%%%%%%%%
\begin{figure}[htb!]
\begin{center}
\includegraphics[scale=0.5]{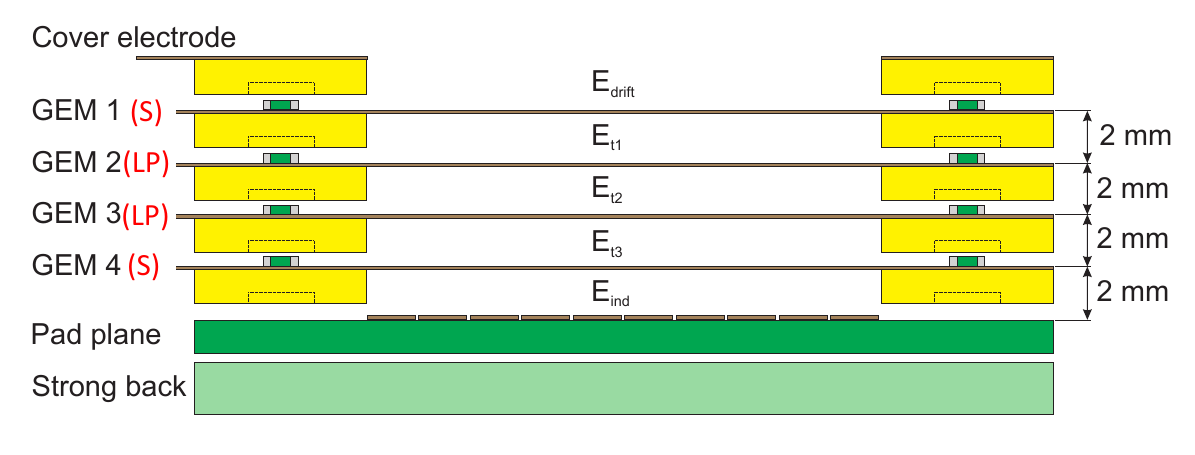}
\caption{\label{four}Schematic exploded cross section of the 4 GEM stack. S - stands for standard GEM foil and LP - stands for larger pitch GEM foil. Each GEM foil is glued onto a 2 mm thick support frame defining the gap. The drift field E$_{drift}$, the transfer fields E$_{ti}$ and the induction field E$_{ind}$ are shown as well \cite{ALICE}.}\label{four}
\end{center}
\end{figure}
%%%%%%%%%%%%%%%%%%%%%%%%%%%%%%%%%%%%%%%%%%%%%%%%%%%%%%%%%%%%%%%%%%%

For Run 3 it is planned to upgrade the ALICE TPC replacing it's MWPC based read-out chambers by a quadruple stacks of GEMs. A schematic representation of such a typical quadruple GEM stack is shown in Figure~\ref{four}. GEM detectors themselves reduce the ion back flow (IBF) without using the GG. GEM also have a high rate capability, and the absence of a long ion signal tail. The basic requirement of this GEM read-out are (1) to operate at a gain of 2000 with Ne/CO$_2$/N$_2$ (90/10/5), (2) the IBF must be $<$ 1\% at an effective gain of 2000 with $\epsilon$ = 20, i.e. 20 ions flowing back into the drift region per incoming electron. (3) In order to retain the performance of the existing system in terms of $\frac{dE}{dx}$ resolution, the local energy resolution of the read-out chambers ($\frac{\sigma_{E}}{E}$) must be $<$12~\% for the 5.9~keV peak of Fe$^{55}$. (4) Stable operation under the LHC conditions. In the upgraded TPC the Neon is chosen because the higher ion mobility as compared to argon. This leads to less space-charge accumulation in the drift field.

%%%%%%%%%%%%%%%%%%%%%%%%%%%%%%%%%%%%%%%%%%%%%%%%%%%%%%%%%%%%%%%%%%%
\begin{figure}[htb!]
\begin{center}
\includegraphics[scale=0.3]{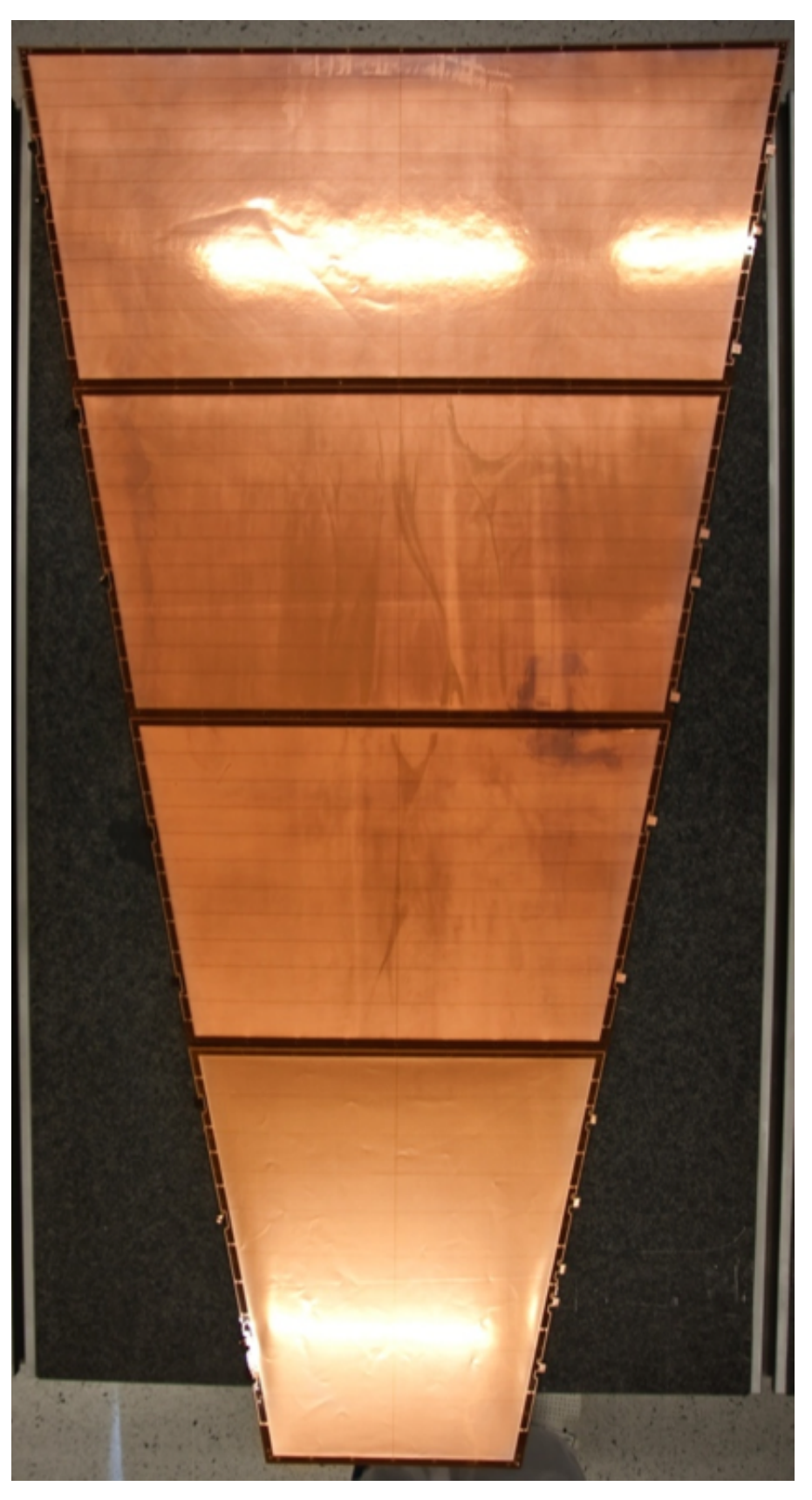}
\caption{\label{roc}One full sector of the ALICE TPC (IROC and OROC) equipped with the four GEM stack set-up (this picture is taken from the Ref~\cite{lippmann}).}\label{roc}
\end{center}
\end{figure}
%%%%%%%%%%%%%%%%%%%%%%%%%%%%%%%%%%%%%%%%%%%%%%%%%%%%%%%%%%%%%%%%%%%

Several full size prototypes have been constructed and tested in beam. The 4-GEM stack set-up needed to equip one full TPC sector is shown in Figure~\ref{roc}. Each sector consists of an Inner Read Out Chamber (IROC, one GEM stack) and an Outer Read Out Chamber (OROC, 3 GEM stacks).

In the fall of 2014 a full-size 4 GEM IROC prototype equipped with the baseline S-LP- LP-S quadruple GEM configuration was built and tested in the T10 beam line of the CERN Proton Synchrotron (PS). The main focus of the PS test beam and the subsequent data analysis was on the study of the particle identification (PID) performance of the detector via the measurement of $\frac{dE}{dx}$. The energy loss $\frac{dE}{dx}$ is determined by the truncated mean of the 70 \% lowest cluster charges along a track. It is calculated from the total charge Q$_{tot}$ of the reconstructed ionisation clusters. Typical energy loss spectra are shown in Figure~\ref{result1}. From these measurements it is verified that the 4-GEM technology offers a $\frac{dE}{dx}$ performance that is compatible with the requirements of the physics programme of the ALICE upgrade \cite{ALICE}. The results are also in very good agreement with the simulation results.

%%%%%%%%%%%%%%%%%%%%%%%%%%%%%%%%%%%%%%%%%%%%%%%%%%%%%%%%%%%%%%%%%%%
\begin{figure}[htb!]
\begin{center}
\includegraphics[scale=0.25]{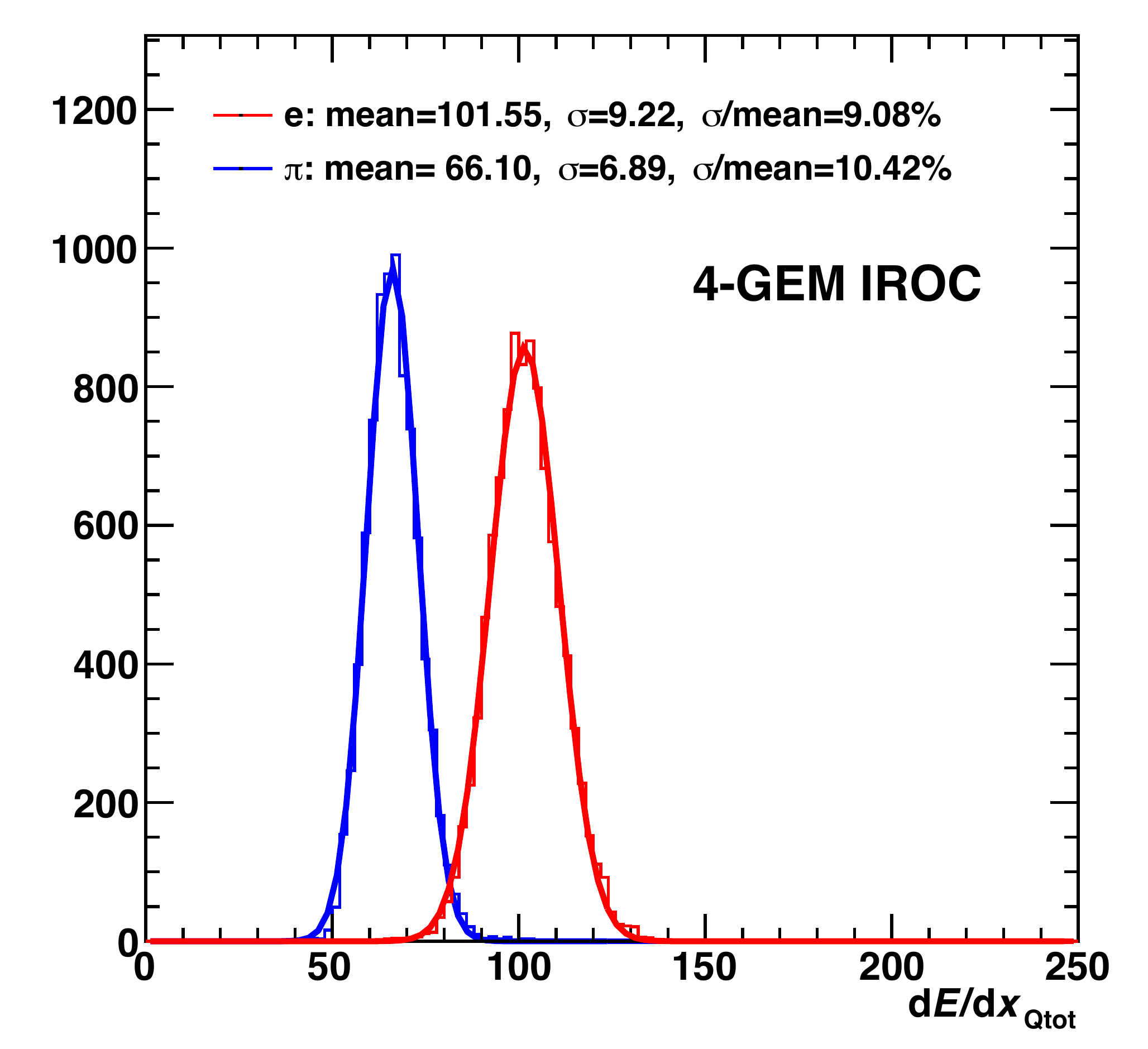}
\caption{\label{result1} The $\frac{dE}{dx}$ distributions of pions (blue) and electrons (red) measured in the 4-GEM IROC at an effective gain of 2000 (this picture is taken from the Ref~\cite{lippmann}).}\label{result1}
\end{center}
\end{figure}
%%%%%%%%%%%%%%%%%%%%%%%%%%%%%%%%%%%%%%%%%%%%%%%%%%%%%%%%%%%%%%%%%%%

The discharge property of the 4-GEM prototype was studied in a dedicated test beam at the CERN Super Proton Synchrotron (SPS) with a gas mixture of Ne/CO$_2$/N$_2$ (90/10/5). The spark probability was measured using showers of hadrons produced by a high-intensity secondary pion beam with a momentum of $\sim$150 GeV/c using a 30~cm thick iron absorber. The average beam intensity was $\sim$ 6 $\times$ 10$^6$ particles per spill of 5 s. During the two-week RD51 test beam campaign, a total of 36 h of dedicated data was collected with the ALICE 4-GEM IROC prototype test set-up. The integral of the continuously recorded, chamber anode current over the whole test beam period gives the total number of accumulated particles (4.7~$\pm$~0.2)~$\times$~10$^{11}$. The discharge performance of the 4-GEM IROC during the test beam campaign is similar to the standard triple GEMs. The discharge performance of the 4-GEM IROC is not likely to cause any damage to the detectors \cite{ALICEtpcadd}. 

In the upgraded ALICE TPC a new Front-End ASIC, called SAMPA, that is mainly based on the present TPC read-out chips, will be used. SAMPA can provide continuous read-out through serial electrical links, programmable conversion gains and peaking times and digital filters for baseline correction (to correct for the common mode effect in the read-out chambers). The average expected data output from the TPC to the online farm for 50 kHz Pb--Pb collisions is 1~TBytes/s.

\section{Indian contribution towards the ALICE TPC upgrade}

An initiative has been taken to perform R\&D of GEM detector prototypes in India for the ALICE TPC upgrade. The triple GEM detector prototype building started. The GEM foils and other components are obtained from CERN. The uniformity of the relative gain was measured on the entire region of the detector. One such detector was tested for long-term stability with a gas mixture of Ar/CO$_2$ of 70/30 and 80/20 volume ratios. The long-term stability test of this detector was performed using a Sr$^{90}$ beta radioactive source and the output anode current was measured with and without source. The temperature (T), relative humidity (RH) and atmospheric pressure (p) were measured and recorded continuously using a data logger developed in-house \cite{SS}. The temperature and pressure (T/p) corrected gain was calculated and plotted as a function of total accumulated charge per unit area. No ageing effect was observed even after operation of the GEM detector for about 700 hours or after an accumulation of charge per unit area of 0.1 mC/cm$^2$ \cite{SB}. A comparison was also made on the long-term stability test results for the Ar/CO$_2$ in 70/30 and 80/20 ratios. For both gas mixtures the T/p corrected anode current is constant at 1 only the fluctuation around 1 is about 8\% for Ar/CO$_2$ in 70/30 whereas it is about 15\% for 80/20 ratio.

\section{Summary and outlook}
A major upgrade of the ALICE experiment will be done in 2018/19. As a part of the ALICE upgrade strategy, it is planned that in order to allow inspection of all Pb--Pb collisions at the full expected LHC heavy ion luminosity of 50~kHz the TPC will be upgraded with 4 GEM read-out chambers. Data from the TPC will be read-out continuously using new front-end electronics called SAMPA ASIC chip and a radiation-hard optical link set. The requirements for the GEM system in terms of the IBF ($<$ 1\%), local energy resolution for the Fe$^{55}$ source ($<$12~\%) and stability against discharges were satisfied in the prototype IROC tests.

In India the long-term stability test has been carried out and is continuing for the triple standard GEM detector with Argon and CO$_2$ based gas mixtures with two different ratios 70/30 and 80/20 and using a Sr$^{90}$ beta source. No ageing has been observed after accumulation charge of around 0.1 mC/cm$^2$. As a future plan a 4-GEM prototype will be built with S-LP-LP-S configuration and the ageing study will be carried out with a 100 mili Curie Fe$^{55}$ source. 

\section{Acknowledgement}
I would like to acknowledge the work done by P. Bhattacharya, B. Mohanty, Rudranarayan Mohanty, Amit Nanda, T. K. Nayak, Rajendra Nath Patra, Sharmili Rudra, Sumanya Sekhar Sahoo, P. K. Sahu, S. Sahu in India. S. Biswas acknowledges the support of DST-SERB Ramanujan Fellowship (D.O. No. SR/S2/RJN-02/2012). This work in India is also funded through DST project no. SB/S2/HEP-022/2013. XII$^{th}$ Plan DAE project titled Experimental High Energy Physics Programme at NISER-ALICE is also acknowledged.


\begin{thebibliography}{99}
\bibitem{lippmann} C. Lippmann, A continuous read-out TPC for the ALICE upgrade, Proceedings of Frontier Detectors for Frontier Physics,13$^{th}$ Pisa meeting on advanced detectors (To be published).
\end{thebibliography}
\end{document}